\documentclass[12pt,english,longbibliography,nofootinbib,superscriptaddress,12pt,sort&compress,showkeys]{revtex4}
\usepackage[latin9]{inputenc}
\usepackage[letterpaper]{geometry}
\usepackage{amsmath}
\usepackage{amssymb}
\usepackage{graphicx}

\makeatletter

\providecommand{\tabularnewline}{\\}

\@ifundefined{textcolor}{}
{%
 \definecolor{BLACK}{gray}{0}
 \definecolor{WHITE}{gray}{1}
 \definecolor{RED}{rgb}{1,0,0}
 \definecolor{GREEN}{rgb}{0,1,0}
 \definecolor{BLUE}{rgb}{0,0,1}
 \definecolor{CYAN}{cmyk}{1,0,0,0}
 \definecolor{MAGENTA}{cmyk}{0,1,0,0}
 \definecolor{YELLOW}{cmyk}{0,0,1,0}
}

\usepackage{indentfirst}
\usepackage{amsfonts}
\usepackage[T1]{fontenc}
\usepackage{ae,aecompl}
\usepackage{sidecap}
\usepackage[section]{placeins}
\usepackage{epsf}
\makeatother

\usepackage{babel}
\date{\today}

\usepackage{chngcntr}

\makeatother

\usepackage{babel}
\begin{document}
\begin{center}
\textbf{\large{}Relationship between grain boundary segregation and
grain boundary diffusion in Cu-Ag alloys} 
\par\end{center}
\author{\noindent R. K. Koju}
\address{Department of Physics and Astronomy, MSN 3F3, George Mason University,
Fairfax, Virginia 22030, USA}
\author{\noindent Y. Mishin}
\address{Department of Physics and Astronomy, MSN 3F3, George Mason University,
Fairfax, Virginia 22030, USA}
\begin{abstract}
While it is known that alloy components can segregate to grain boundaries
(GBs), and that the atomic mobility in GBs greatly exceeds the atomic
mobility in the lattice, little is known about the effect of GB segregation
on GB diffusion. Atomistic computer simulations offer a means of gaining
insights into the segregation-diffusion relationship by computing
the GB diffusion coefficients of the alloy components as a function
of their segregated amounts. In such simulations, thermodynamically
equilibrium GB segregation is prepared by a semi-grand canonical Monte
Carlo method, followed by calculation of the diffusion coefficients
of all alloy components by molecular dynamics. As a demonstration,
the proposed methodology is applied to a GB is the Cu-Ag system. The
GB diffusivities obtained exhibit non-trivial composition dependencies
that can be explained by site blocking, site competition, and the
onset of GB disordering due to the premelting effect.
\end{abstract}
\maketitle

\section{Introduction}

Solute segregation to grain boundaries (GBs) can affect many mechanical,
thermodynamic and kinetic properties of materials \citep{Balluffi95}.
Once at GBs, the solute atoms can diffuse through the material much
faster than they would by regular, lattice diffusion mechanisms \citep{Kaur95}.
The accelerated atomic transport along GBs, often referred to as ``short-circuit''
diffusion, can control the kinetics of processes such as creep deformation
\citep{Coble,Farghalli01,Mishin-2013}, phase precipitation \citep{Christian_2002,Porter-Easterling},
as well as complex kinetic phenomena such as dynamic strain aging
\citep{Schwink97,Curtin06a}. 

It is well-established that the self-diffusion and solute diffusion
coefficients in GBs can exceed the lattice diffusion coefficients
by many orders of magnitude, especially at low homologous temperatures
\citep{Kaur95}. What remains poorly understood is how the amount
of GB segregation can affect the rate of GB diffusion. For example,
in a binary alloy A-B one must consider GB diffusion coefficients
of both the solute component B ($D_{B}$) as well as the host component
A ($D_{A}$). Several questions arise. For example, as the amount
of GB segregation of B increases, do the diffusion coefficients $D_{A}$
and $D_{B}$ both increase, both decrease, or can show opposing trends?
Which physical factors control the effect of segregation on GB diffusion?
Can the segregation-diffusion relation change with temperature and/or
alloy composition? How does the frequently occurring disordering of
the GB structure at high temperatures can affect the segregation-diffusion
relation? 

To our knowledge, these questions remain largely open. Answering them
by experiment is not impossible in principle but is hampered by technical
obstacles. One of them being that, to keep track of both diffusion
coefficients ($D_{A}$ and $D_{B}$) in the same GB, a co-diffusion
experiment is required with concurrent monitoring of the segregated
amounts. The experiment would then have to be repeated for a set of
alloy compositions and/or temperatures. Such experiments are technically
challenging and, to our knowledge, have not been performed so far.
Another challenge is related to the fact that GB diffusion experiments
are predominantly carried out at relatively high temperatures at which
a significant fraction of the atoms diffusing along the GB leaks into
the surrounding lattice regions \citep{Kaur95}. Under such conditions,
called the type-B kinetic regime, one can only extract from the experiment
the triple product $sD_{A,B}\delta$, $s$ being the segregation factor
and $\delta$ the GB width. Separate determination of the GB diffusion
coefficients $D_{A,B}$ requires specially designed low-temperature
experiments conducted in the so-called type-C regime. C-regime measurements
are much more difficult and have only been performed for a small number
of systems \citep{Kaur95,Mishin97e,Mishin99f,Herzig:2003aa,Divinski:2004aa,Divinski01a,Divinski2012}.
Such systems do not include alloys with a varied chemical composition.
Furthermore, only the solute diffusivity $D_{B}$ has been measured
in the C-regime.

Given the experimental challenges mentioned above, a meaningful alternative
approach is offered by atomistic computer simulations. It has recently
been demonstrated that GB diffusion coefficients can be reliably computed
in pure metals as well as dilute binary alloys (in the latter case,
for solute diffusion only) \citep{Sorensen00,Suzuki05a,Frolov09c,Frolov2013a}.
This methodology can serve as a starting point from which to launch
a systematic study of the effect of GB segregation on GB diffusion
of \emph{both} chemical components in binary, and in the future multicomponent,
alloy systems.

The goal of this paper is to initiate work in the outlined direction
by performing a series of simulations of GB segregation and GB diffusion
in Cu-rich Cu-Ag solid solutions chosen here as a model system. The
Cu-Ag system has the advantage of exhibiting a limited solid solubility
of the two elements and a strong GB segregation trend. Its choice
also puts us on a familiar ground since much information has already
been obtained for this system in previous work \citep{Williams06,Frolov:2015aa,Frolov:2016aa,Hickman:2016aa}.
In particular, a reliable interatomic potential is available \citep{Williams06},
and the phase diagram predicted by this potential has been accurately
computed \citep{Williams06,Hickman:2016aa}. GBs in Cu have been studied
extensively \citep{Suzuki03a,Suzuki03b,Suzuki05a,Suzuki05b,Cahn06a,Cahn06b,Mishin07a,Spearot07,Tschopp07b,Tschopp07a,Tschopp07,Frolov09c,Frolov2012b,Frolov2013,Fensin2012,Frolov2013a,Frolov:2015aa,Frolov:2015ab}.
One typical GB was chosen here as an example, with the intent of extending
this work to a larger set of boundaries in the future. We perform
a detailed study of Ag GB segregation in a wide temperature-composition
domain of the Cu-Ag system, followed by a similarly detailed study
of GB diffusion of both Ag and Cu and its correlation with the segregation
behavior.

\section{Methodology}

\noindent Atomic interactions in the Cu-Ag system were modeled using
an embedded atom potential \citep{Williams06} that accurately reproduces
a large number of physical properties of both Cu and Ag. The potential
was fitted to first-principles energies of Cu-Ag compounds and predicts
the Cu-Ag phase diagram in reasonable agreement with experiment. Molecular
dynamics (MD) simulations were performed using the Large-scale Atomic/Molecular
Massively Parallel Simulator (LAMMPS) \citep{Plimpton95}. The Monte-Carlo
(MC) simulations  utilized the parallel MC code ParaGrandMC developed
by V. Yamakov at NASA \citep{ParaGrandMC,Purja-Pun:2015aa,Yamakov:2015aa}.

\begin{figure}[ht]
\noindent \centering{}\includegraphics[width=0.5\textwidth]{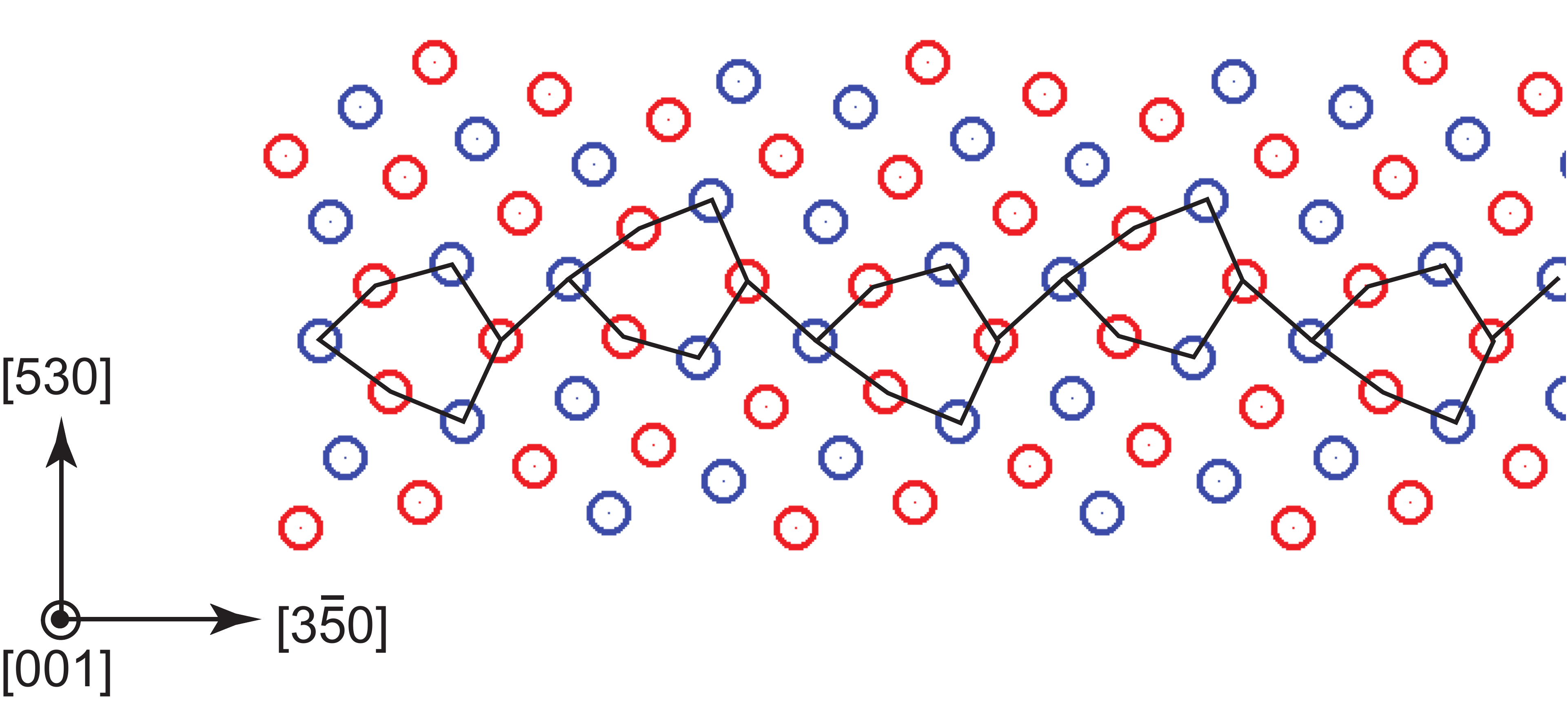}
\caption{Atomic structure of the $\Sigma17(530)[001]$ symmetrical tilt GB
in Cu. The red and blue circles represent the atoms in alternating
(002) planes normal to the {[}001{]} tilt axis. The structural units
are outlined. \label{fig:GB17530}}
\end{figure}

As a representative high-angle GB, we chose the symmetrical tilt $\Sigma17(530)[001]$
GB with the misorientation angle of $61.93^{\circ}$. Here, $\Sigma$
is the reciprocal density of coincident sites, {[}001{]} is the tilt
axis, and $(530)$ is the GB plane. The boundary was created in a
rectangular periodic simulation block whose edges were, respectively,
parallel to the tilt axis ($x$-direction), normal to the tilt axis
($y$-direction), and normal to the GB plane ($z$-direction). The
block had the approximate dimensions of $10.54\times10.48\times21.13$
nm$^{3}$ and contained $1.97\times10^{5}$ atoms. The ground-state
structure of the GB in pure Cu was obtained by the $\gamma$-surface
method \citep{Mishin98a,Suzuki03a,Suzuki03b}. The structure consists
of identical kite-shaped structural units arranged in a zig-zag array
as shown in Fig. \ref{fig:GB17530}. The rows of structural units
running parallel to the tilt axis can be interpreted as closely spaced
edge dislocations forming the GB core. The same structure of this
GB was previously obtained in Cu \citep{Suzuki05a,Cahn06b} and Ni
\citep{Sun:2020aa}. The GB energy was found to be 856 mJ/m$^{2}$
in agreement with previous reports \citep{Suzuki05a,Cahn06b}.

A prescribed amount of Ag was introduced into Cu by semi-grand canonical
MC simulations implemented at a chosen temperature $T$ and a fixed
value of the chemical potential difference between Ag and Cu. The
trial moves of the MC process included random displacements of randomly
selected atoms with a random re-assignment of their chemical species
to either Ag or Cu. The trial move additionally included random changes
in the dimensions of the simulation block with rescaling of the atomic
coordinates to achieve zero pressure conditions in all three directions.
The trial move was accepted or rejected by the Metropolis algorithm.
The simulation produced a thermodynamically equilibrium distribution
of Ag atoms in the GB region and inside the grains for the targeted
alloy composition. The simulations covered the temperature range between
600 K and 1100 K, with the alloy compositions varying from pure Cu
to the solidus line. 

The amount of Ag segregation was quantified by the excess number of
Ag atoms per unit GB area at a fixed total number of atoms:
\begin{equation}
[N_{\textnormal{Ag}}]=N_{\textnormal{Ag}}-N\dfrac{N_{\textnormal{Ag}}^{\prime}}{N^{\prime}},\label{eq:1}
\end{equation}
where $N_{\textnormal{Ag}}$ and $N_{\textnormal{Ag}}^{\prime}$ are
the numbers of Ag atoms per unit area in two regions with and without
the GB, respectively, and $N$ and $N^{\prime}$ are the respective
total numbers of Cu and Ag atoms. Both regions were large enough to
include both the GB and the interiors of the grains.

The degree of structural disorder in the GB was measured by the layer-averaged
structure factor $S(\mathbf{k})$. The simulation block was divided
into 0.1 nm thin layers parallel to the GB plane and numbered
by index $i$. The structure factor corresponding to layer $i$ is
defined by 
\begin{equation}
S_{i}(\mathbf{k})=\frac{1}{N_{i}}\sqrt{\sum_{j=1}^{N_{i}}\cos^{2}(\mathbf{k}\cdot\mathbf{r}_{j})+\sum_{j=1}^{N_{i}}\sin^{2}(\mathbf{k}\cdot\mathbf{r}_{j})},
\end{equation}
where $\mathbf{k}=2\pi[2/a,0,0]$ is the chosen reciprocal lattice
vector, $\mathbf{r}_{j}$ is the position of atom $j$ within the
layer $i$, $a$ is the cubic lattice parameter, and $N_{i}$ is the
total number of atoms in the layer. The structure factor so defined
equals one in the perfect lattice at 0 K, has a value $S_{\infty}(\mathbf{k})<1$
in the lattice at finite temperatures, and turns to zero in the liquid
phase. It is expected to exhibit a local minimum at the GB position
due to the local disorder. The value of the structure factor relative
to the lattice value, $\varphi(z_{i})=S_{i}(\mathbf{k})-S_{\infty}(\mathbf{k})$,
is defined as the order parameter at position $z_{i}=\lambda i$ in
the GB region ($\lambda$ being the layer thickness). Furthermore,
the width $w$ of the order parameter minimum can be taken as the
structural width of the GB. Specifically, $w$ was defined as twice
the standard deviation of the Gaussian fitted to the order parameter
profile $\varphi(z_{i})$ across the GB. Knowing the GB width, the
Ag concentration in the GB can be found by averaging the atomic fraction
of Ag over the layer of width $w$ centered at the Gaussian peak.
This concentration provides a complementary measure of the GB segregation
in addition to $[N_{\textnormal{Ag}}]$.

GB diffusion coefficients were computed from MD simulations performed
on GBs pre-equilibrated by MC simulations. First, the potential energy
peak across the current GB position was constructed by averaging the
potential energy over thin layers parallel to the boundary plane.
The peak width was typically around 1 nm or larger. Mean-square atomic
displacements, $\left\langle x^{2}\right\rangle $ and $\left\langle y^{2}\right\rangle $,
parallel to the GB plane were computed as functions of time for both
Ag and Cu atoms. The calculations only included atoms within a 1 nm
thick window centered at the boundary position. The mean-square displacements
were monitored over a period of time $\Delta t$ ranging from 24 ns
to 60 ns, depending on the alloy composition and temperature. The
GB diffusion coefficients of Ag and Cu in both directions were obtained
from the Einstein relations $D_{x}=\left\langle x^{2}\right\rangle /2\Delta t$
and $D_{y}=\left\langle y^{2}\right\rangle /2\Delta t$, respectively.
Due to the structural anisotropy of the GB, the diffusion coefficients
parallel ($D_{x}$) and normal ($D_{y}$) to the tilt axis are generally
different. To account for slight variations in the GB position with
time due to thermal fluctuations, the 1 nm layer in which the mean-square
displacements were calculated was periodically re-centered to the
current GB position identified with the potential energy peak.

\section{Results}

\subsection{Grain boundary segregation}

\begin{figure}[ht]
\noindent \centering{}\includegraphics[width=0.6\textwidth]{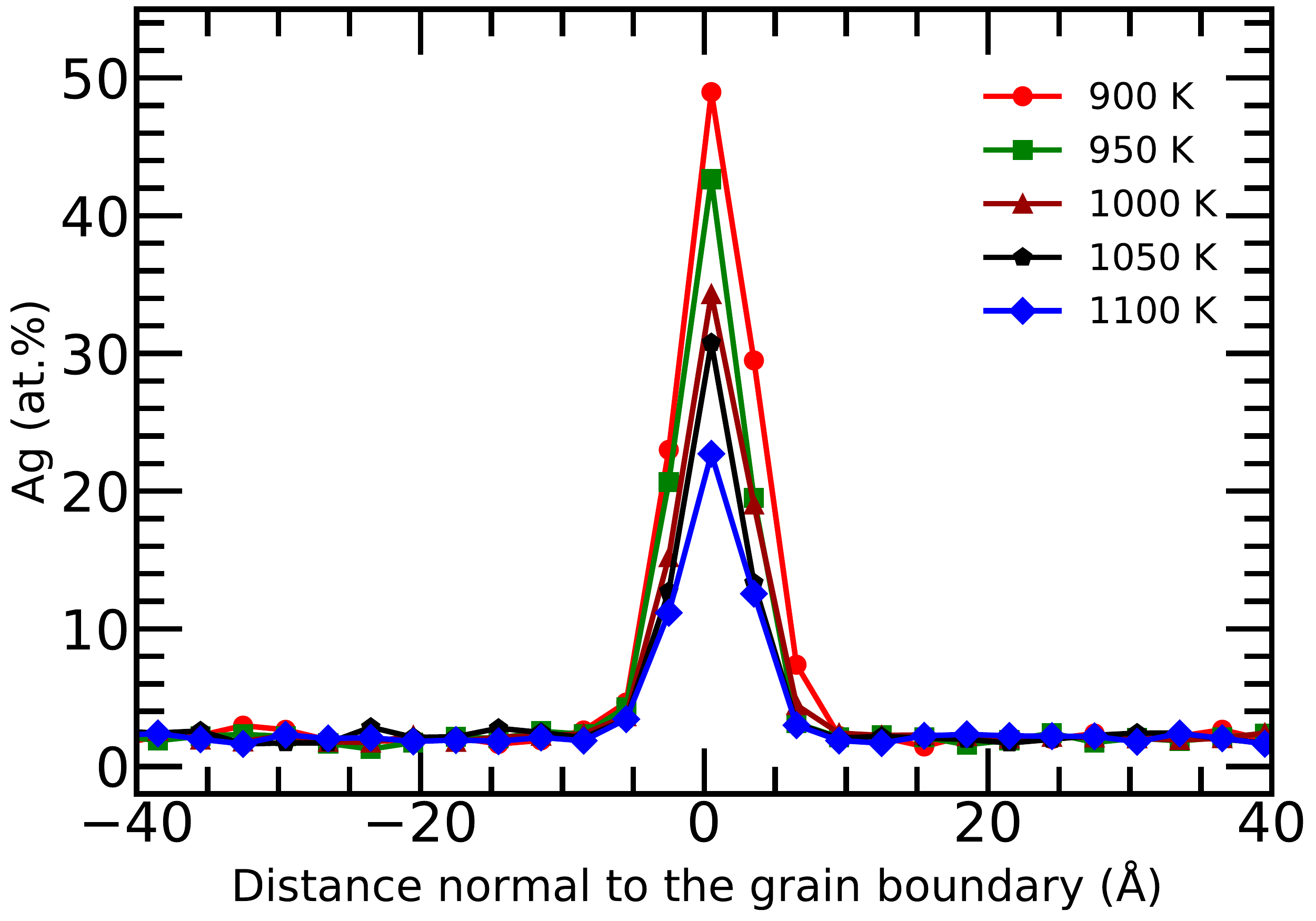}
\caption{Ag GB segregation profiles in the Cu-2\,at.\%Ag alloy at various
temperatures. \label{fig:Segreg_profiles}}
\end{figure}

Fig.~\ref{fig:Segreg_profiles} illustrates typical equilibrium segregation
profiles in the Cu-2\,at.\% alloy at various temperatures. The profiles
were obtained by averaging the atomic fraction of Ag over thin layers
parallel to the GB and then averaging over multiple snapshots saved
during the MC simulations. Note that the segregation peak grows higher
with decreasing temperature and broadens with increasing composition.
As will be discussed below, the width of the segregation zone drastically
increases near the solidus line as the GB undergoes the premelting
transformation.

\begin{figure}[ht]
\noindent \centering{}\includegraphics[width=0.57\textwidth]{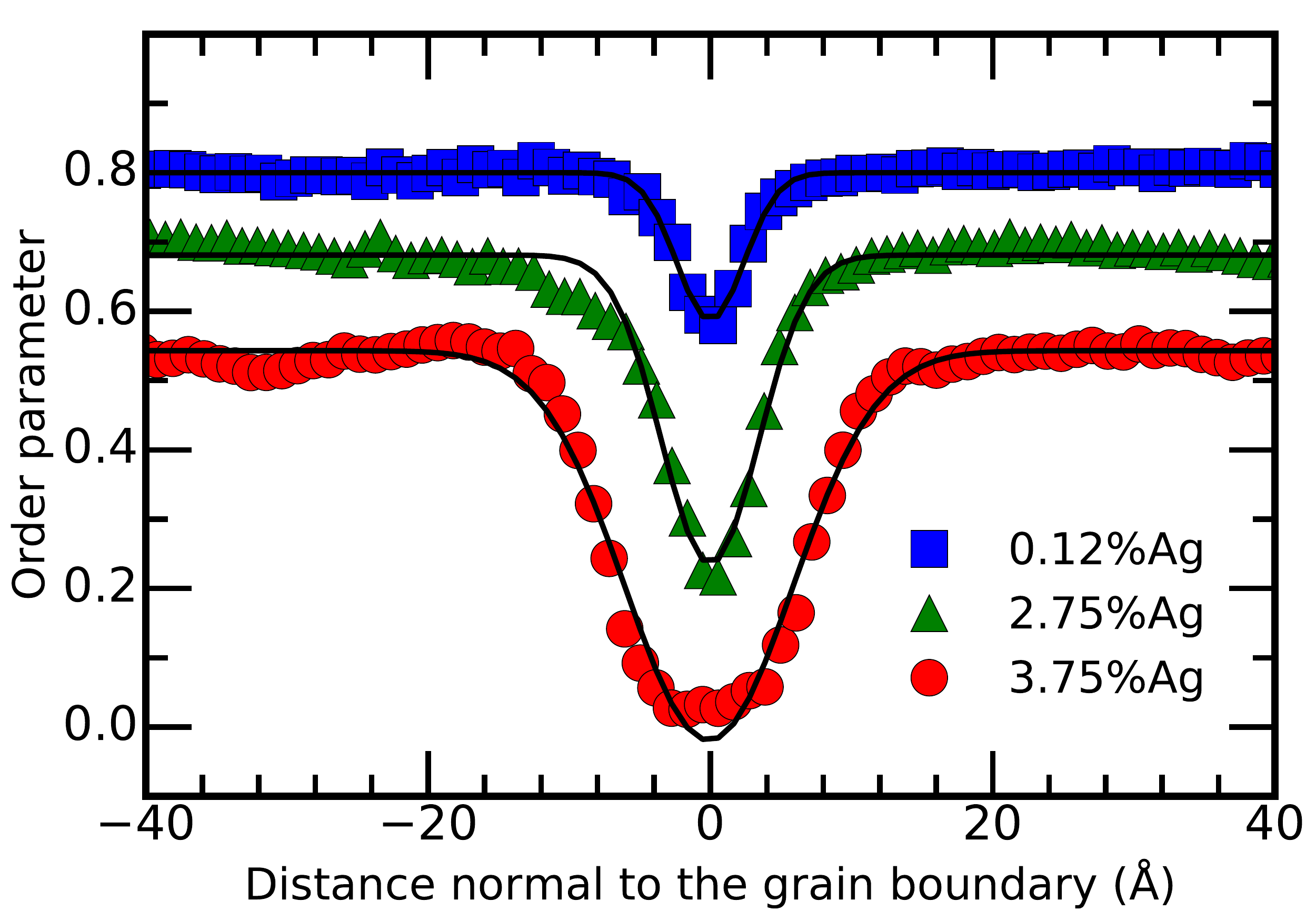}
\caption{Representative profiles of the order parameter $\varphi(z)$ across
the GB for three alloy compositions at the temperature of 1100 K.
The curves represent Gaussian fits of the local minimum occurring
at the GB position. \label{fig:Order_profiles}}
\end{figure}

\begin{figure}[ht]
\noindent \begin{centering}
(a)\includegraphics[width=0.55\textwidth]{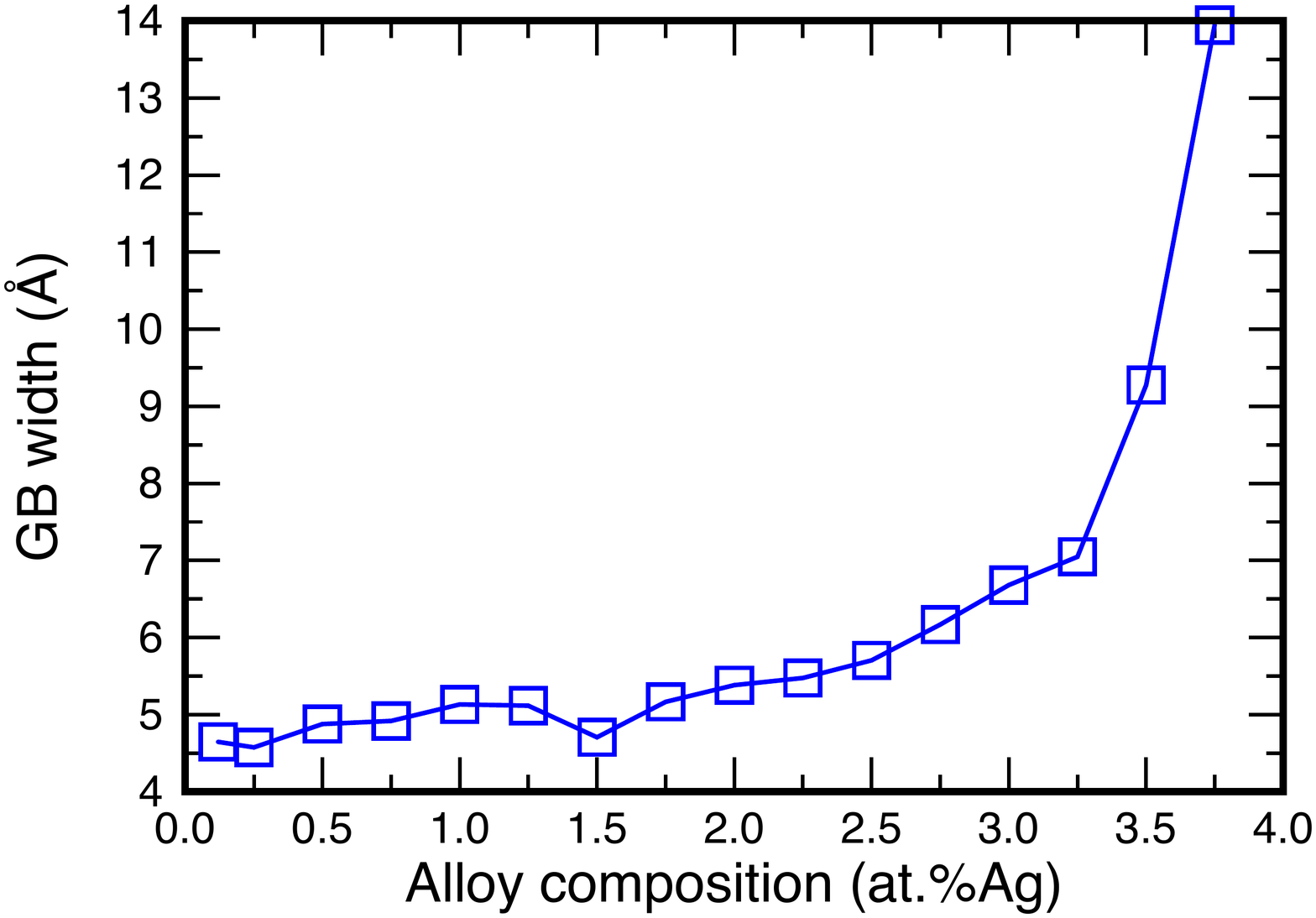} 
\par\end{centering}
\bigskip{}
\bigskip{}

\noindent \centering{}(b)\includegraphics[width=0.55\textwidth]{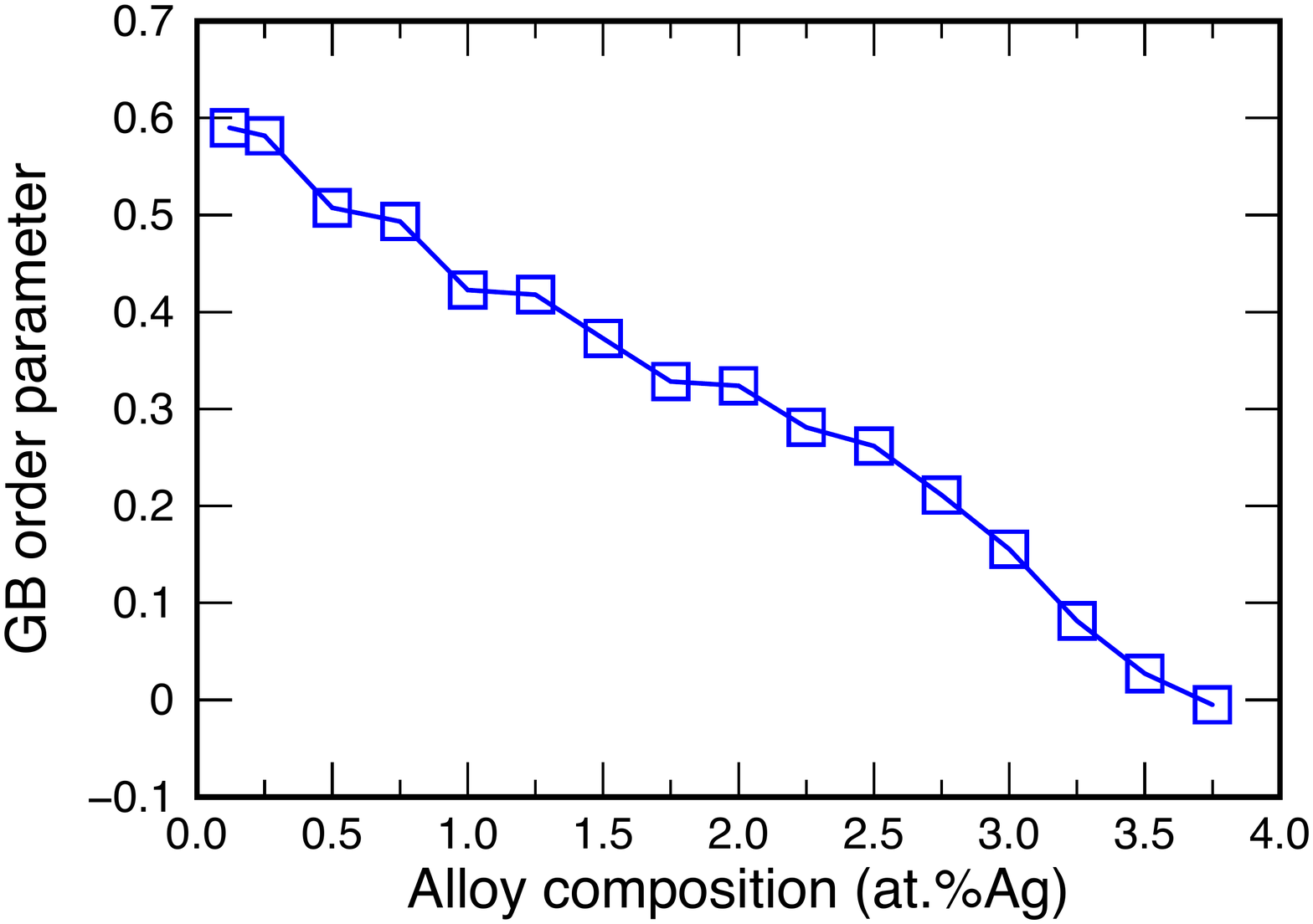}
\caption{(a) GB width $w$ and (b) GB order parameter $\varphi(0)$ as functions
of alloy composition at the temperature of 1100 K. Note that $w$
diverges to infinity while $\varphi(0)$ tends to zero at the solidus
composition of about 4 at.\%Ag. \label{fig:Order_width}}
\end{figure}

Representative order parameter profiles $\varphi(z)$ are shown in
Fig.~\ref{fig:Order_profiles}. At a fixed temperature (1100 K in
this case), the minimum becomes deeper as Ag concentration increases,
indicating the accumulation of structural disorder in the GB core.
As the alloy composition approaches the solidus line, the order parameter
in the GB tends to zero ($\varphi(0)\rightarrow0$), while the GB
width $w$ rapidly increases and eventually spreads across the entire
simulation block (Fig.~\ref{fig:Order_width}). This behavior is
a clear manifestation of GB melting and a sign that the alloy composition
has reached the solidus line at the given temperature. 

\begin{figure}[ht]
\noindent \begin{centering}
(a)\includegraphics[width=0.65\textwidth]{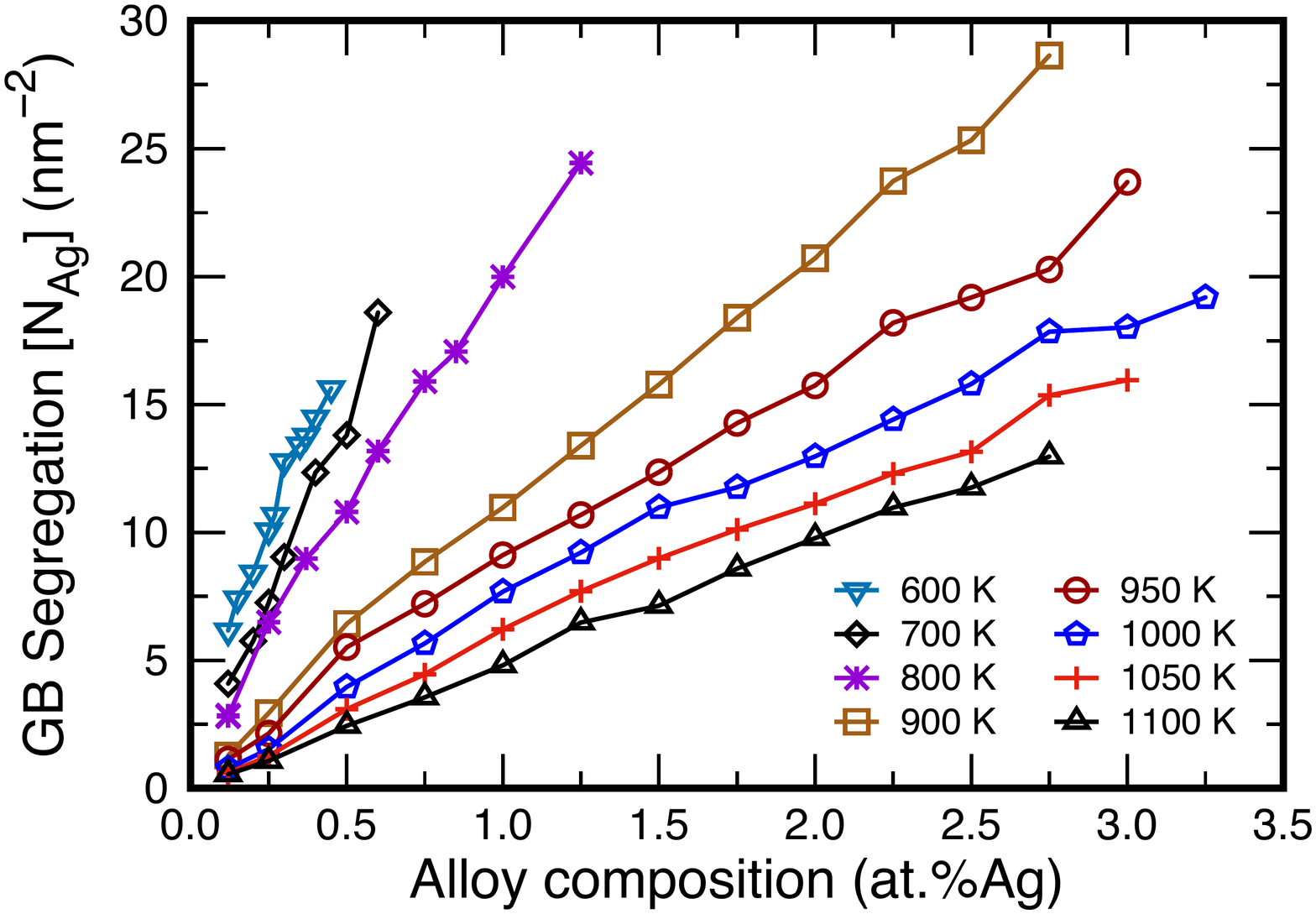}
\par\end{centering}
\bigskip{}
\bigskip{}

\noindent \centering{}(b)\includegraphics[width=0.65\textwidth]{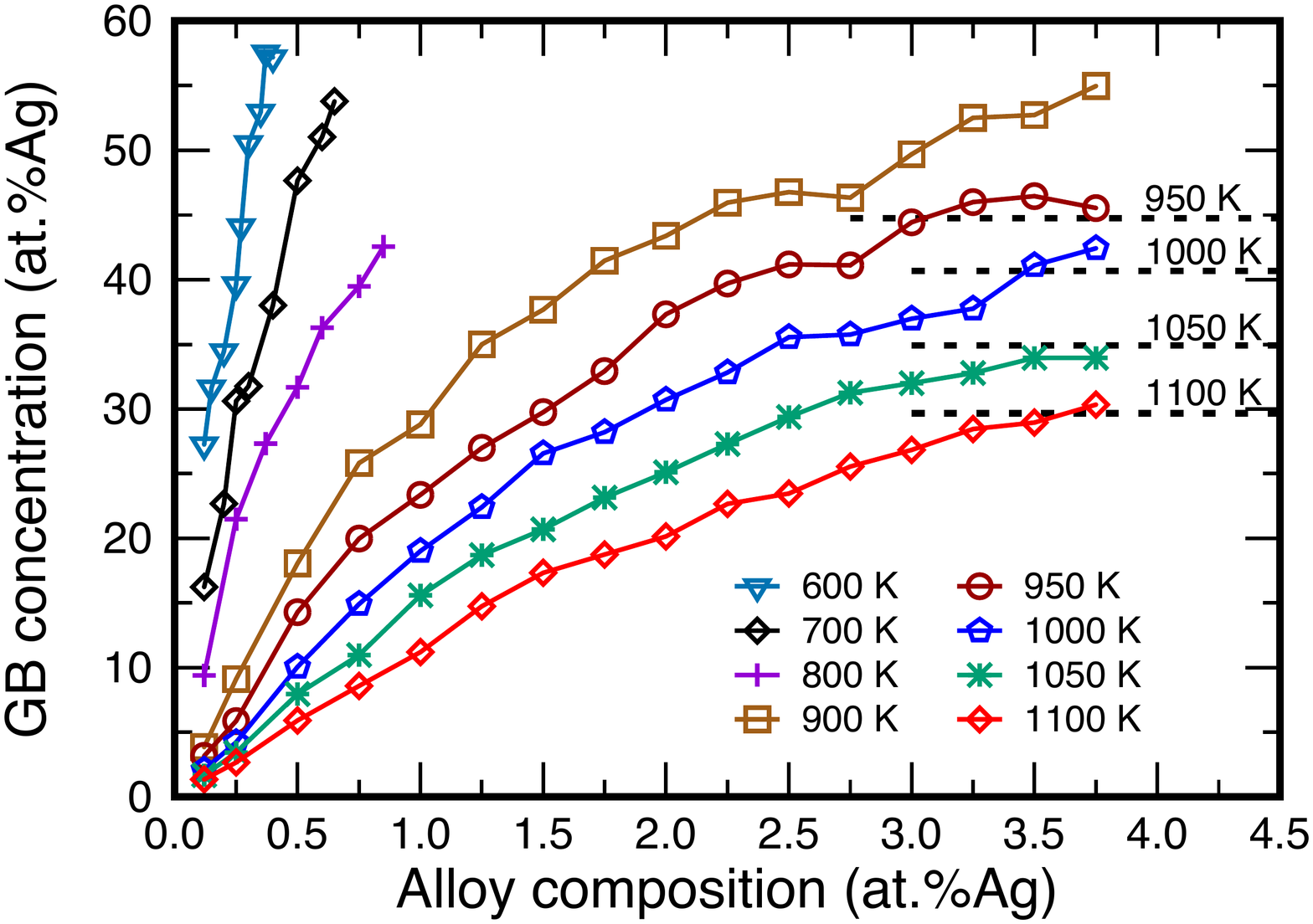}
\caption{(a) Amount of Ag GB segregation $[N_{\textnormal{Ag}}]$ and (b) GB
composition (atomic percentage of Ag atoms) as functions of alloy
composition at different temperatures. Each curve ends at the solidus
line on the phase diagram. In (b), the dashed lines represent the
liquidus compositions obtained from the phase diagram at temperatures
$\protect\geq950$ K. \label{fig:Isotherms}}
\end{figure}

Isotherms of GB segregation are plotted in Fig.~\ref{fig:Isotherms}
using two measures of segregation: the total segregated amount $[N_{\textnormal{Ag}}]$
(number of excess Ag atoms per unit area) and the chemical composition
(at.\%Ag) within the GB core. Both segregation parameters increase,
in a nonlinear manner, with increase in the alloy concentration and
decrease in temperature. Larger $[N_{\textnormal{Ag}}]$ values result
from both the increase in the GB concentration and the GB broadening
effect (Fig.~\ref{fig:Isotherms}a). By contrast, the isotherms shown
in Fig.~\ref{fig:Isotherms}b capture the behavior of the GB composition
alone. Note that, at temperatures above the eutectic temperature predicted
by the interatomic potential ($T_{E}=935$ K \citep{Williams06}),
the GB composition reaches the liquidus composition on the computed
phase diagram \citep{Williams06}. Thus, at temperatures above $T_{E}$,
the GB transforms into a liquid layer of the liquidus composition
when the grain composition approaches the solidus line. GB melting
behavior in the Cu-Ag system was also noted in previous simulation
studies \citep{Williams09,Hickman:2016aa}.

Distribution of the segregated Ag atoms inside the GB was examined
in detail using the OVITO visualization software \citep{Stukowski2010a}.
In dilute compositions, the GB remained highly ordered and the segregated
Ag atoms substituted for the host Cu atoms at particular positions
within the GB structural units (Fig.~\ref{fig:Segreg_snapshots}a).
As the alloy concentration increased, the GB structure grew increasingly
disordered (Fig.~\ref{fig:Segreg_snapshots}b) until the structural
units could no longer be distinguished (Fig.~\ref{fig:Segreg_snapshots}c).
We emphasize that this disordering effect was entirely caused by the
Ag segregation. In pure Cu, the GB structure remained well-ordered
until high temperatures approaching the Cu melting point (1326 K \citep{Mishin01}).

\subsection{Grain boundary diffusion}

The GB diffusion coefficients were computed at temperatures and alloy
compositions lying within the Cu-based solid solution domain on the
Cu-Ag phase diagram. For the chosen GB, the diffusion coefficients
parallel ($D_{x}$) and normal ($D_{y}$) to the tilt axis were found
to be nearly equal. Thus, only the average values $D=(D_{x}+D_{y})/2$
are reported below.

\begin{figure}[ht]
\noindent \begin{centering}
(a)\includegraphics[width=0.75\textwidth]{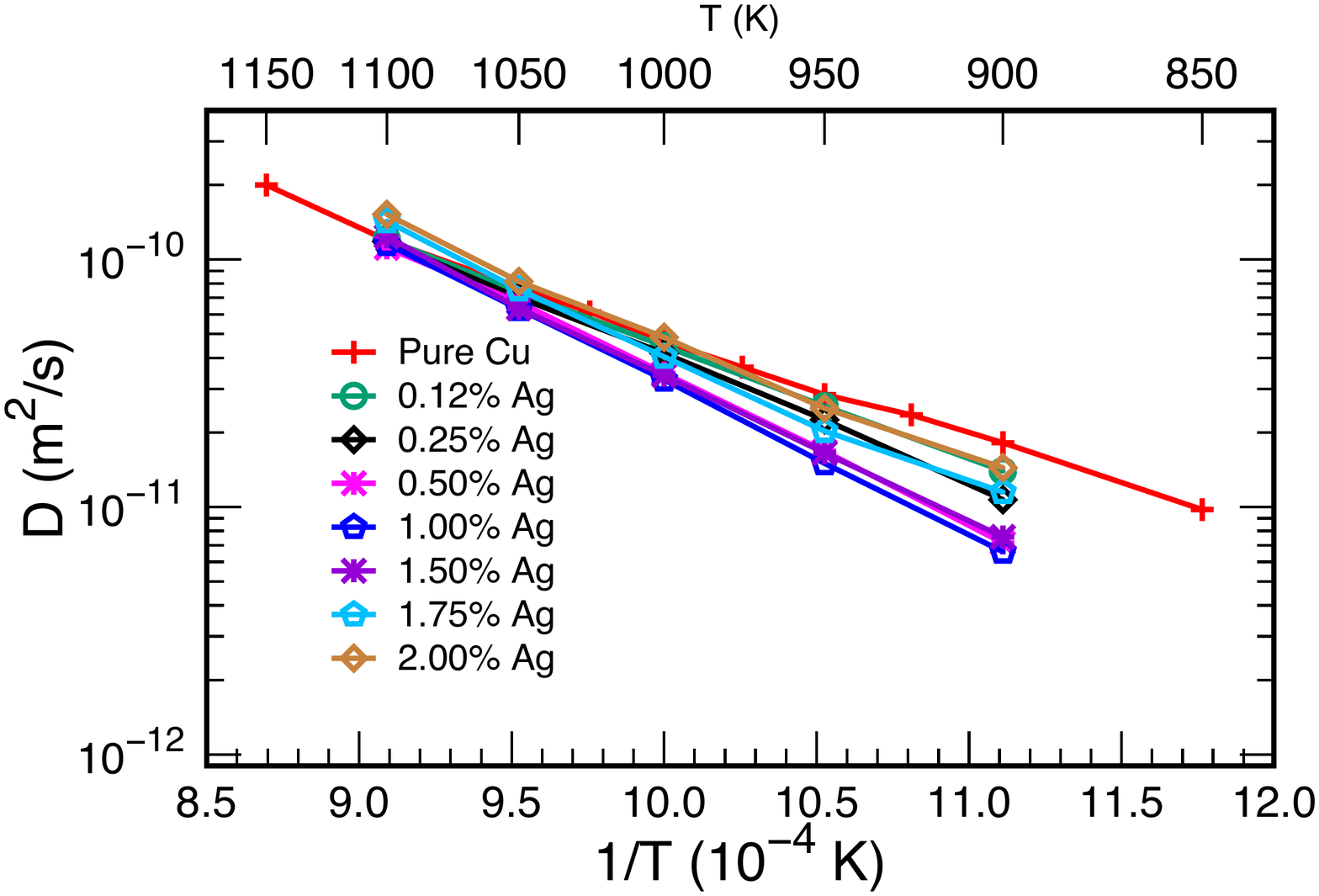}
\par\end{centering}
\bigskip{}
\bigskip{}

\noindent \centering{}(b)\includegraphics[width=0.75\textwidth]{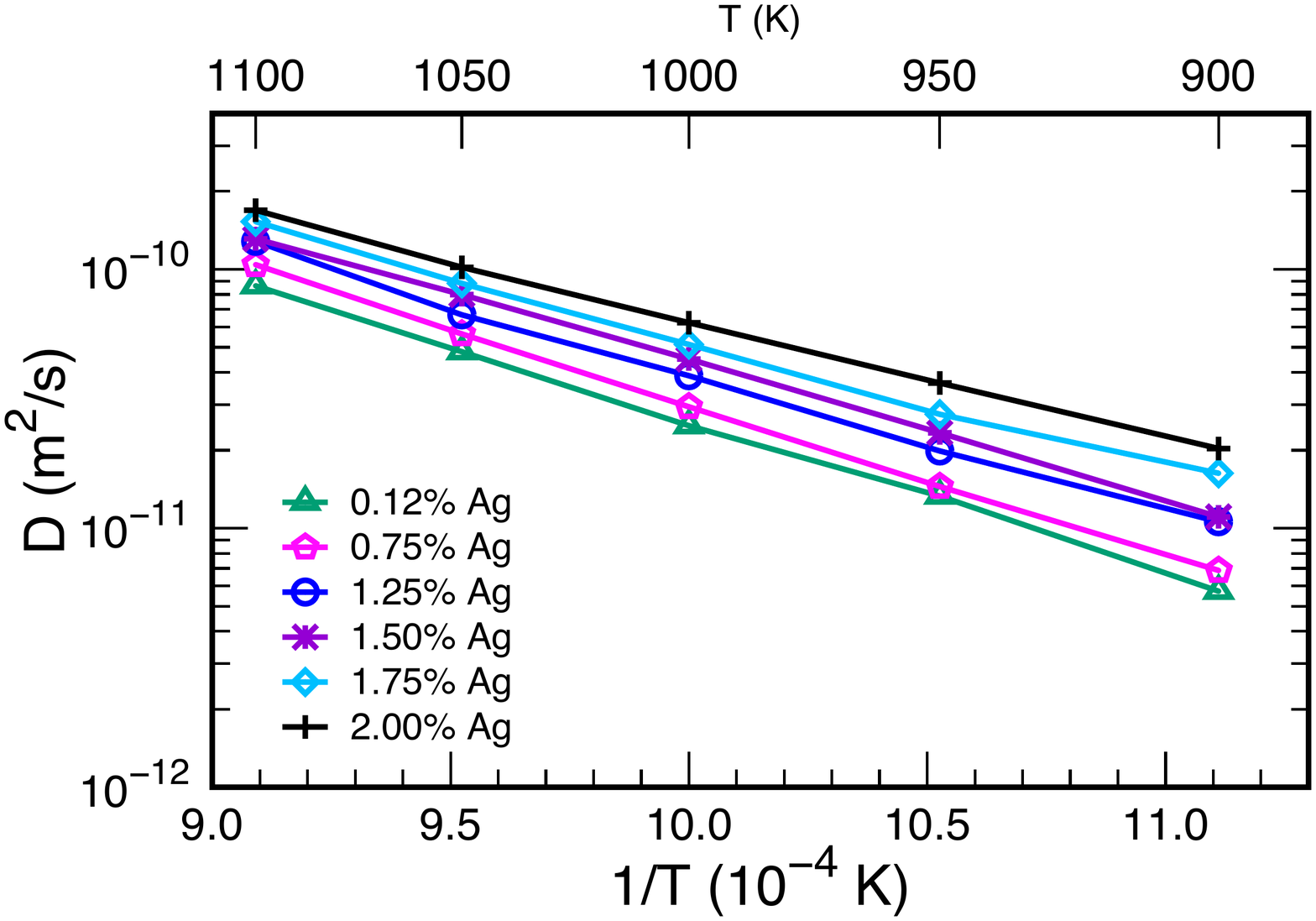}
\caption{Arrhenius diagrams of GB diffusion coefficients of (a) Cu and (b)
Ag in Cu-Ag alloys with different chemical compositions. \label{fig:Arrhenius}}
\end{figure}

The GB diffusion coefficients obtained are summarized on the Arrhenius
diagrams, $\log D$ versus $1/T$, shown Fig.~\ref{fig:Arrhenius}a
(Cu diffusion) and Fig.~\ref{fig:Arrhenius}b (Ag diffusion). The
alloy compositions are limited to 2 at.\%Ag to avoid close proximity
of the solidus line. While diffusion in highly premelted GBs representing
liquid layers could also be measured, the results would not be relevant
to the segregation-diffusion relationship pursued in this work.

\begin{figure}[ht]
\noindent \begin{centering}
(a)\includegraphics[width=0.65\textwidth]{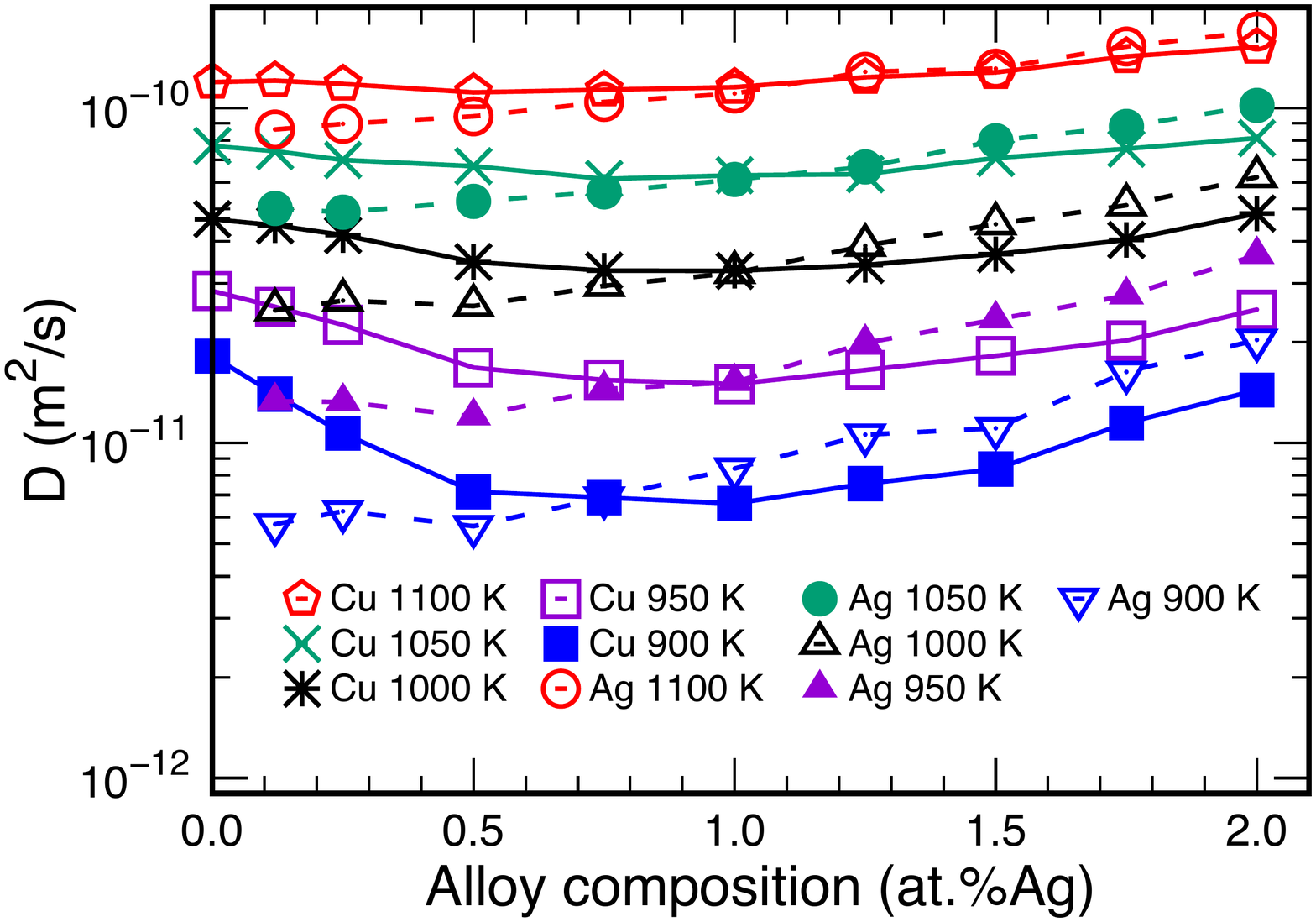}
\par\end{centering}
\bigskip{}
\bigskip{}

\noindent \centering{}(b)\includegraphics[width=0.65\textwidth]{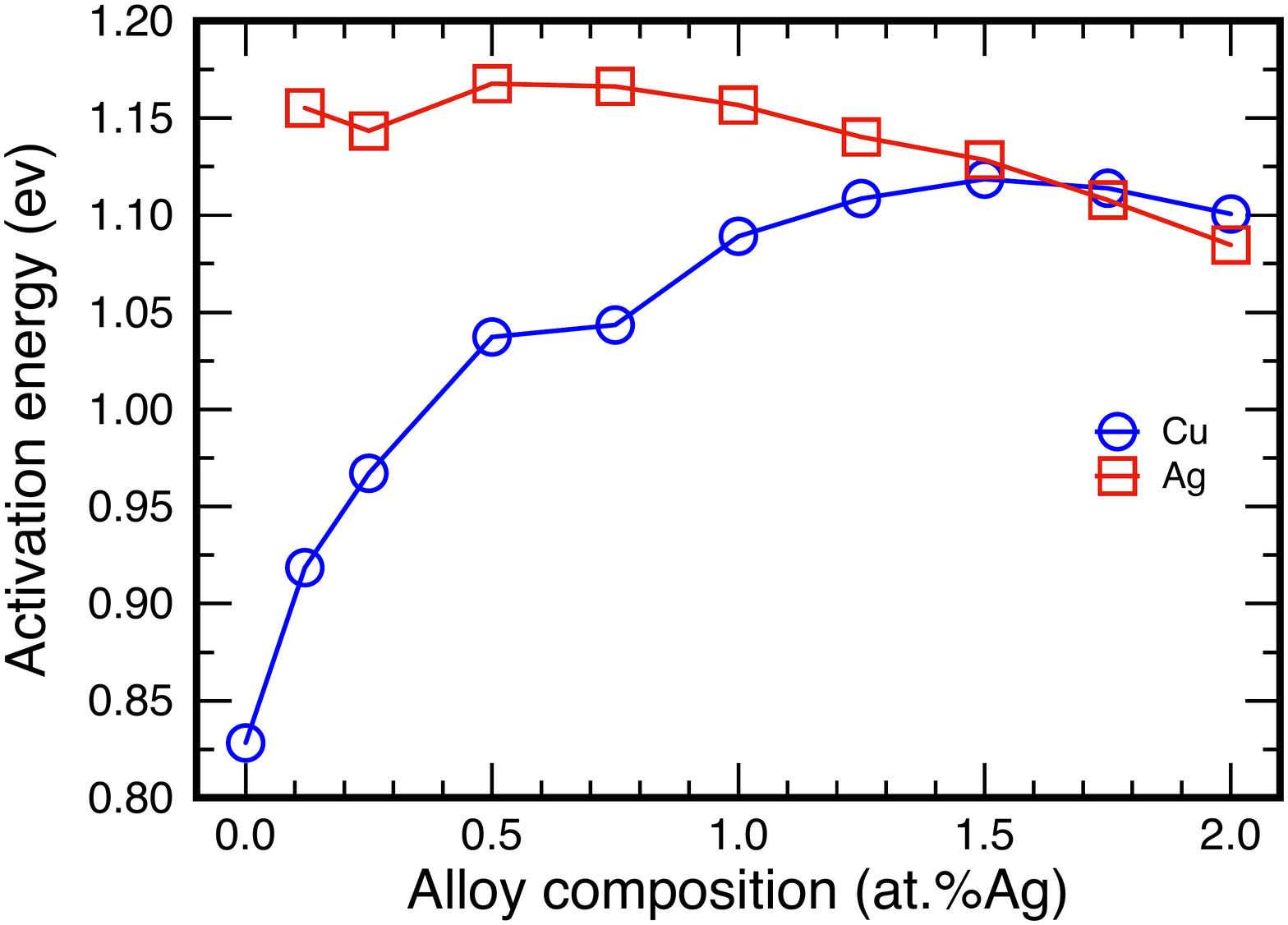}
\caption{(a) GB diffusion coefficients of Cu and Ag as a function of alloy
composition at different temperatures. The data points are connected
by solid (Cu) and dashed (Ag) lines as a guide to the eye. (b) Activation
energy of Cu and Ag GB diffusion as a function of alloy composition.
\label{fig:GB-diffusion}}
\end{figure}

The diffusion coefficients in Fig.~\ref{fig:Arrhenius} reasonably
follow the Arrhenius relation
\begin{equation}
D=D_{0}\exp\left(-\dfrac{E}{kT}\right)\label{eq:Arrhenius-eqn}
\end{equation}
at all temperatures. The plots demonstrate that the diffusion coefficients
of both components depend on the alloy composition. To display the
composition dependence more clearly, we plot the diffusion coefficients
as a function of at.\%Ag in Fig.~\ref{fig:GB-diffusion}a. Two trends
are obvious:
\begin{itemize}
\item Ag atoms diffuse in the GB slower than the host Cu atoms at low concentrations
but faster at higher concentrations. The crossover occurs at about
1 at.\%Ag. 
\item While the Ag diffusion coefficients increase with Ag concentration
monotonically, the Cu diffusion coefficients display a non-monotonic
composition dependence, with a local minimum occurring at about 1
at.\%Ag.
\end{itemize}
Note that the diffusion coefficients are shown in Fig.~\ref{fig:GB-diffusion}a
on the logarithmic scale, meaning that the trends described are quite
significant. The following explanation of these trends can be proposed.
At low temperatures, the Ag atoms tend to segregate to particular
GB sites offering the largest segregation energy. Due to this energetic
preference, the Ag atoms spend most of the time occupying such favorable
sites. They are reluctant to jump to alternate sites (i.e., against
the driving force) to participate in the diffusion process, which
results in slower diffusion rates. As additional Ag atoms segregate
to the GB, they are forced to occupy less favorable (higher energy)
sites and are more likely to contribute to the diffusion flux. In
other words, the trapping effect weakens and Ag diffusion accelerates
as the alloy concentration increases. At the same time, the Cu atoms
diffuse slower with the addition of Ag due to the site blocking effect:
the less mobile Ag atoms disrupt the fast diffusion pathways for Cu
diffusion within the GB structure. As a result, the Ag and Cu diffusivities
display opposite trends, converging toward each other as clearly observed
in Fig.~\ref{fig:GB-diffusion}a.

\begin{figure}[ht]
\noindent \begin{centering}
(a)\includegraphics[width=0.4\textwidth]{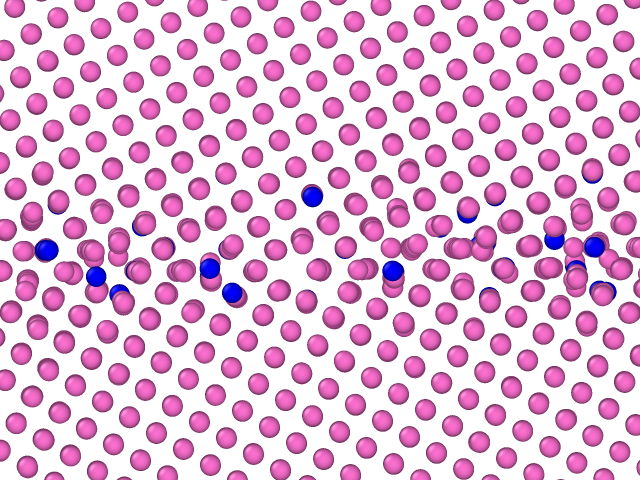}
\par\end{centering}
\bigskip{}
\bigskip{}

\noindent \begin{centering}
(b)\includegraphics[width=0.4\textwidth]{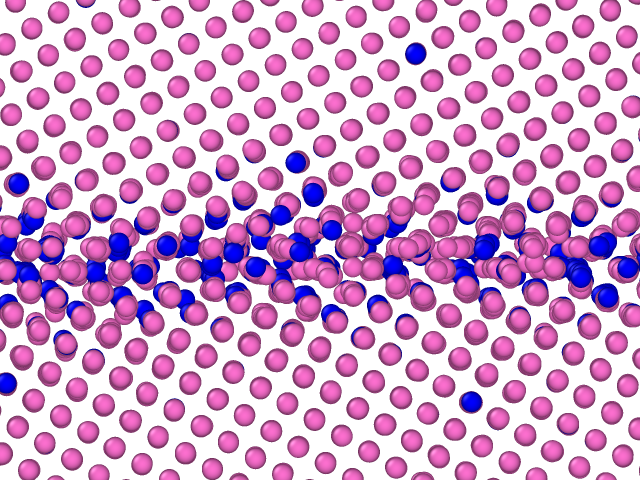} 
\par\end{centering}
\noindent \begin{centering}
\bigskip{}
\bigskip{}
\par\end{centering}
\noindent \centering{}(c)\includegraphics[width=0.4\textwidth]{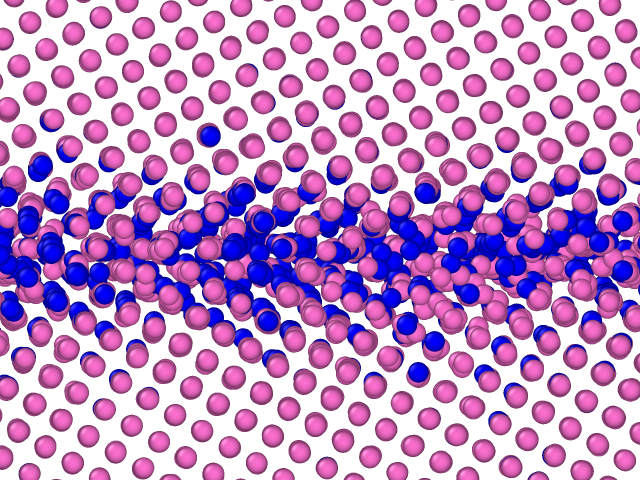}
\caption{Distribution of Ag atoms in the GB at the alloy compositions of (1)
Cu-0.12\,at.\%, (b) Cu-1\,at.\%Ag, and (c) Cu-2\,at.\%Ag at the
temperature of 900 K. The Ag and Cu atoms are shown in blue and pink,
respectively. Note the accumulation of GB disorder with increase in
the GB segregation. \label{fig:Segreg_snapshots}}
\end{figure}

This explanation only applies as long as the GB maintains an ordered
structure with well-defined structural units offering distinct types
of segregation site. This is certainly true for dilute alloy compositions
as illustrated in Fig.~\ref{fig:Segreg_snapshots}a. At higher Ag
concentrations when the GB develops a significant disorder (Fig.~\ref{fig:Segreg_snapshots}b)
and eventually transforms into a liquid-like state (Fig.~\ref{fig:Segreg_snapshots}c),
the situation changes. Diffusion in disordered GBs is governed by
different atomic mechanisms from those in ordered structures \citep{Suzuki05a,Mishin:2015ac},
hence a change in the diffusion trend with composition can be expected.
This change can explain the crossover of the Ag and Cu diffusivities
and the existence of a local minimum of the Cu GB diffusivity at about
1 at.\%Ag. This is the approximate composition at which the GB disordering
commences at the temperatures studied here (Fig.~\ref{fig:Segreg_snapshots}b).

The crossover effect also manifests itself in the composition dependence
of the activation energy $E$ of GB diffusion appearing in Eq.(\ref{eq:Arrhenius-eqn}).
While Ag GB diffusion is characterized by a higher activation energy
in comparison with Cu below about 1 at.\%Ag, the two activation energies
converge to each other in more concentrated alloys in which the GB
loses the ordered structure (Fig.~\ref{fig:GB-diffusion}b).

For validation of our methodology, we can compare the activation energies
computed in this work with experimental data available in the literature
(Table \ref{tab:Activation-energy}). For GB self-diffusion in Cu,
only data for polycrystals is available \citep{Surholt97a}. The reported
activation energy varies between $E=$ 0.751 eV and 0.878 eV, depending
on the chemical purity of the material \citep{Surholt97a}. Our calculations
predict $E=$ 0.828 eV, which we consider a good agreement given that
the polycrystalline value of $E$ represents an average over many
GB types. For Ag GB diffusion, the experiments give $E=$ 1.126 eV
(in pure Cu \citep{Divinski01a}) and 1.128 eV (in Cu-0.2\,at.\%Ag
\citep{Divinski2012}), in both cases for polycrystalline samples.
The closest chemical compositions studied in this work are Cu-0.12\,at.\%Ag
and Cu-0.25\,at.\%Ag. The respective activation energies, 0.918 eV
and 0.967 eV, compare well with the experiment considering that they
were obtained for one particular GB. Another piece of experimental
information comes from a recent study of Ag diffusion in a Cu bicrystal
with the $\Sigma5(310)[001]$ GB \citep{Divinski2012}. Even though
this boundary is different from ours and is considered special, the
experimental activation energy (0.983 eV or 1.067 eV, depending on
the diffusion direction) is close to our results for the $\Sigma17(530)[001]$
boundary in the dilute limit. Thus, the comparison with experiment
is very encouraging and lends confidence to the simulation results
reported in this paper.

\section{Conclusions}

The goal of this work was to demonstrate that it is now possible to
probe the effect of GB segregation on GB diffusion of \emph{both}
the solute and solvent components in alloys by means of atomistic
computer simulations. The methodology proposed combines MC simulations
to create an equilibration GB segregation with MD simulations to extract
the GB diffusion coefficients. A reliable interatomic potential is
required, and the relevant part of the phase diagram must be known
or computed.

As an example, we have studied diffusion in a representative GB in
the Cu-Ag system in the temperature-composition domain of Cu-based
solid solutions. Our results indicate that the GB diffusivities of
the solute (Ag) and solvent (Cu) atoms can exhibit quite different
and non-trivial composition/temperature dependencies. They can correlate
with each other, anti-correlate, cross, or have local minima. These
behaviors reflect intricate interplays between different diffusion
mechanisms and physical effects, such as site blocking and site competition.
One factor that is more crucial in alloys than it is in elemental
solids is the disordering of the GB structure. When the alloy composition
and/or temperature approach the solidus line on the phase diagram,
GBs can become atomically disordered at relatively low temperatures,
eventually transforming to a liquid film \citep{Suzuki05a,Williams09,Mishin09c,Hickman:2016aa}.
This disordering is fueled by GB segregation and can drastically alter
the GB diffusion mechanisms and thus the segregation-diffusion relationship
in comparison with ordered GB structures prevailing in elemental solids
and/or solid dilute solutions. 

This work was performed on one particular GB in one binary system.
Future studies in the proposed direction may include larger GB sets,
multicomponent systems, and a more detailed analysis of the underlying
diffusion mechanisms.

\vspace{0.15in}

\textbf{Acknowledgement:} This work was supported by the National
Science Foundation, Division of Materials Research, under Award No.\,1708314.


\newpage\clearpage{}

\begin{table}

\caption{Activation energies of GB diffusion obtained by the present simulations
in comparison with experimental data from the literature \citep{Herzig:2003aa,Divinski2012,Divinski01a,Surholt97a}.
For Ag GB diffusion in Cu-Ag alloys, two chemical compositions are
included as closest to the experimental composition of Cu-0.2\,at.\%Ag
\citep{Herzig:2003aa}. $^{a}$ parallel to the tilt axis, $^{b}$
normal to the tilt axis.\label{tab:Activation-energy}}
\bigskip{}

\begin{tabular}{|c|c|c|c|c|}
\hline 
Diffusing atoms & Material & GB type & Activation energy $E$ (eV) & Reference\tabularnewline
\hline 
\hline 
Cu & 99.999\% Cu & Polycrystal & 0.878 & \citep{Surholt97a}\tabularnewline
\hline 
Cu & 99.9998\% Cu & Polycrystal & 0.751 & \citep{Surholt97a}\tabularnewline
\hline 
Cu & Cu & $\Sigma17(530)[001]$ & 0.828 & This work\tabularnewline
\hline 
Ag & Cu & Polycrystal & 1.126 & \citep{Divinski01a}\tabularnewline
\hline 
Ag & Cu-0.2\,at.\%Ag & Polycrystal & 1.128 & \citep{Herzig:2003aa}\tabularnewline
\hline 
Ag & Cu & $\Sigma5(310)[001]$ & 0.983$^{a}$; 1.067$^{b}$ & \citep{Divinski2012}\tabularnewline
\hline 
Ag & Cu-0.12\,at.\%Ag & $\Sigma17(530)[001]$ & 0.918 & This work\tabularnewline
\hline 
Ag & Cu-0.25\,at.\%Ag & $\Sigma17(530)[001]$ & 0.967 & This work\tabularnewline
\hline 
\end{tabular}
\end{table}
\newpage\clearpage{}

\end{document}